# Curvature sensing by vision and touch

Birgitta Dresp-Langley

*ICube UMR 7357 CNRS - University of Strasbourg*

**Abstract**

Brain representations of curvature may be formed on the basis of either vision or touch. Experimental and theoretical work by the author and her colleagues has shown that the processing underlying such representations directly depends on specific two-dimensional geometric properties of the curved object, and on the symmetry of curvature. Virtual representations of curves with mirror symmetry were displayed in 2D on a computer screen to sighted observers for visual scaling. For tactile (hapticà scaling, the physical counterparts of these curves were placed in the two hands of sighted observers, who were blindfolded during the sensing experiment, and of congenitally blind observers, who never had any visual experience. All results clearly show that curvature, whether haptically or visually sensed, is statistically linked to the same curve properties. Sensation is expressed psychophysically as a power function of any symmetrical curve's *aspect ratio*, a scale invariant geometric property of physical objects. The results of the author's work support biologically motivated models of sensory integration for curvature processing. They also promote the idea of a universal power law for adaptive brain control and balancing of motor responses to environmental stimuli across sensory modalities.

**Introduction**

Manipulation of visual objects with the two hands and the visual and tactile integration of shape information play an important role in action planning and motor control [1, 2]. This is particularly important in spatial perception by the blind who never had visual experience (congenitally blind people), but who are nonetheless perfectly capable of understanding the physical environments which surround them, and of forming exact representations of complex spatial geometry. Understanding how the blind compensate through touch for lack of visual data is relevant to understanding how visual and tactile sensory integration is interactively programmed in the human brain, and has implications for medical robotics, i.e. the study of motor planning and control [3] as well as clinical neurology, i.e. the study and treatment of neurological disorders such as spatial neglect [4, 5] or tactile allodynia [6]. Physical as well as perceptual models are needed to extend our knowledge on how visual and tactile brain representations function and interact.

Since the early theories of active perception proposed by Gibson [12, 13], who was among the first to suggest that visual and tactile representations share common functional properties, neuroscientists have been looking for neural correlates of perception for action. The cross-modal brain processing of signals mediated by vision and touch involves cortical neurons with non-classic receptive field structures. Such have been functionally identified in the primate brain [7, 8, 9, 10, 11], and these receptive fields belong to multimodal neurons corresponding to specific perception-action spaces. The sensorial pathways for visual and tactile processing are independent at early stages of encoding. At later stages, a coupling of signals must take place in the brain to enable sensory coordination. This inevitably involves feed-back signals from motor responses. It is likely that power laws, known to govern the integration of chemosensory signals in modality specific pathways, reflect the generic mechanism that enables the invariant control and balance of brain activities related to the processing of signals from different sensory pathways. Yet, model hypotheses backed by consistent psychophysical data are still lacking.

Curvature is a mathematical property of the physical world that is of considerable importance in complex system science. Curvature guides physical, chemical, and biological processes, such as protein folding, membrane binding, and other biophysical transformations [14]. The representation and cognition of curvature starts at the biochemical level of living organisms capable of sensing this property in their near or distant physical environments [15]. Curvature is also a perceptual property extracted from physical objects (Figure 1) to form veridical representations in highly developed organisms such as the brain. The ability of processing curvature (the spatial contrast of a curve against its background) visually can be linked to receptive field properties of visual cortical neurons (Figure 2), and it was found that such ability is highly developed in humans. The spatial precision with which man can distinguish a curve from a straight line reflects only a fifth of the physical spacing between two neighboring visual receptors, and a tenth of the smallest receptive field centre of ganglion cells found in the primate retina [16, 17]. Among the geometric properties of curves exploited by the brain to achieve such an astonishing performance are the distance between the line or chord that joins the two ends of a curve and the parallel line tangent to that curve, the sagitta [18] and the ratio between the sagitta and the length of the chord, also called aspect ratio [19]. Visual images of curves with the same aspect ratio but varying sagitta produce identical detection thresholds, suggesting that the visual processing of curves may well be independent of scale. The aspect ratio of a curve is a scale-invariant parameter which conveys a global representation of the spatial extent or area covered by the curve, while the sagitta provides a strictly local cue to the point of maximum curvature. A mathematical model for planar curve generation based on vertically and horizontally oriented ellipses was found to produce the most reliable psychometric functions for visually mediated estimates of curvature as a function of the aspect ratio of a large number of planar curves, presented in random order on a computer screen to human observers [20]. The stimuli used in these visual experiments were designed to reflect properties of real world objects that can be manipulated with the hands.

The most parsimonious mathematical definition of curvature relates to circles and ellipses. In terms of geometry, curves derived from circles and ellipses share certain properties, the circle being a particular case of the ellipse. Also, the choice of elliptic curves for studying perceptual mechanisms is biologically motivated given their symmetry and the observation that, in the real world, the curvature of the contours of natural objects corresponds to a wide range of symmetrical shapes [21-27] with the Euclidean properties of ellipses. For the purpose of this study, we used projective geometry to generate curves from ellipses, by affinity with concentric circles, as will be explained in the following section.

**Material and Methods**

Computer generated virtual curves and their real-world counterparts were presented to human observers for psychophysical scaling in two separate experiments. One experiment consisted of a randomly presented sequence of virtual curves presented as visual stimuli on a computer screen to observers with normal vision. In the other experiment, the real-world counterparts of these curves were presented manually, in random order and according to the same procedure, to blindfolded and congenitally blind observers. In both tasks, observers had to rate the perceived magnitude of each stimulus on a standard psychophysical scale [21, 22] with values ranging from 0 for no curvature perceived to 10 for maximum curvature perceived. This classic psychophysical scaling procedure has proven a reliable tool for studying perceptual sensations and their internal representation. Psychophysical scaling aims at linking psychological and physiological mechanisms to the physical or mathematical properties of the outside world and was introduced at the beginning of last century by eminent medical scientists such as Fechner and Wundt. The method is now widely applied in contemporary medical research and clinical testing.

*Mathematical properties of the curves*

The curves were derived from planar ellipses, generated in AUTOCAD on the basis of projective geometry, through transformation by affinity with concentric circles (Figure 1 top), a procedure frequently used in digital rendering and design. To understand how ellipses are obtained in such a way, it is useful to recall some of the properties of concentric circles, which share the same centre. In the Cartesian plane, the principal circle with centre 0 ($C_{0,a}$) is defined in terms of

$$R^2\ (C_{0,a}) = (x)^2 + (y)^2$$

where $R$ is the radius of the circle, and $x$ and $y$ the two-dimensional spatial coordinates of the points falling on its perimeter. The second, concentric circle is obtained from the first one by

$$R^2\ (C_{0,b}) = (x+\delta x)^2 + (y+\delta y)^2$$

or

$$R^2\ (C_{0,b}) = (x-\delta x)^2 + (y-\delta y)^2$$

Ellipses as projected images of concentric circles (Figure 1 top) may be defined in terms of

$$(x,y) = (bx, ay)$$

of the principal circle C(0,a) and

$$(x,y) = ((a/b)x, y)$$

of the secondary circle C(0,b). This transform is sometimes referred to as a particular case of Newton's transform [27, 28]. In the Cartesian plane, the ellipse (*E*) is defined in terms of

$$E = x^2/a^2 + y^2/b^2 = 1$$

axes *a* and *b* being the axes of symmetry intersecting at its center (Figure 3 top). The larger axis of the two (*a*) is referred to as the major, and the smaller (*b*) as the minor. The majors and the minors are directly linked to the sagitta, or maximum height (*H*), and the chordlength, or width (*W*), of the curves that were derived from the ellipses here (Figure 3 bottom), which were cut in half at one of the axes of symmetry to generate curves with varying orientation in the plane (upward or downward), varying height (*H*), and varying width (*W*).

*Virtual curves for visual presentation*

Curves for visual presentation were generated in AUTOCAD and stored as individual images in an image library for the experiments, which were run on an IBM Pentium III equipped with a standard colour screen. The size of a single pixel on the screen corresponded to 0.0025 cm. The physical curve parameters of maximum height (*H*) and width (*W*) translate into degrees of visual angle (*DegV*), which reflects the size of a given parameter as it is perceived by the human eye as a function of the distance of the observer from the object viewed, using the linear transform

$$DegV = 2 arctan(S/2D)$$

where *S* is the physical stimulus parameter (in centimetres) and *D* the viewing distance (also in centimetres) or distance of the observer's head from the screen, held constant at 120 centimetres in this experiment. All stimuli were presented foveally, bearing in mind that the diameter of foveal vision is limited to approximately 5 degrees of visual angle.

*Real-world curves for tactile exploration*

The real-world counterparts of the curves for the tactile experiments were designed by the same group of design students. Ellipses with the same height and width properties as those generated in AUTOCAD to build a visual library of curve images presented for presentation on a computer screen were drawn on cardboard with compass and ruler following a method for drawing ellipses manually (the "gardener's ellipse method", described on http://www.mathopenref.com. The curves, made of flexible plastic-coated wire cable were bent by hand and matched visually to the images of the half ellipses. The cable material was flexible, but rigid enough for a given curve once formed to retain its shape reliably under gentle manipulation with the fingers of the two hands.

*Participants*

Six men and four women, between 25 and 32 years old and equipped with normal vision, participated in the visual experiments. Five women and three men, between 26 and 48 years old and equipped with normal vision but blindfolded for the purpose of the experiment, and two congenitally blind observers, both men, participated in the tactile experiments . Each of the individuals only participated in one of the two experiments, which were conducted in accordance with the Declaration of Helsinki (1964). None of the participants was aware of the aims of the study.

*Procedures*

Eleven curves with positive curvature (upward orientation) and eleven curves with negative curvature (downward orientation) in the two-dimensional plane were presented on a computer screen in the visual experiment. The stimuli were presented in random order in 22 successive trials. Each curve was presented once for two seconds in a session. Curve orientation (upward or downward) also varied randomly. All curves were presented at high contrast, defined by a contour segment of the thickness of a single pixel on the screen with a luminance of 40cd/m$^2$.

The luminance of the screen background on which the curves were presented was 2cd/m$^2$. Observers were seated comfortably in a semi-dark room, with their heads resting on a head-and-chin rest at a distance of 120 centimetres from the computer screen. Their hands rested on a desk with a computer keyboard, which they had to use to indicate the perceived magnitude of curvature for each stimulus by typing a number from zero to ten on the computer keyboard. Typing the "enter" key then triggered presentation of the next stimulus in the visual experiment. Observers were instructed that they were going to view a series of curves, one at a time, and were asked to produce a number between 0 and 10 that was to reflect the intensity (magnitude) of curvature they spontaneously perceive when a given curve comes up on the screen. A straight line was shown at the beginning of each individual session, solely to make sure that the observer spontaneously typed "0" on the computer keyboard and had, indeed, understood the instruction to scale curvature. Psychophysical scaling does not require giving lower and upper limits of a physical stimulus to a healthy adult human observer. Once fully developed, the non-pathological human brain is capable of reliably scaling any stimulus on the basis of internally represented (learnt) upper and lower limits [21]. In the tactile experiment, observers were instructed that they were going to explore a series of curved cables, one at a time, with the fingers of their two hands. They were asked to produce a number between 0 and 10 that was to reflect the intensity (magnitude) of curvature they perceive when a given curve is placed in their hands, and to explore each curve very gently without pulling or bending the wire cables. An experimenter was present to make sure that this instruction was adhered to and that none of the observers deformed a cable. A perfectly straight wire cable was placed in the two hands of an observer at the beginning of the trials to make sure that each of them spontaneously replied "zero curvature" and that the instruction to scale curvature had been understood. The stimuli were given to the observers, in random order, one at a time. Each of the eleven curves was presented to each observer twice, once pointing upward and once pointing downward, generating a sequence of 22 successive trials as in the visual experiment.

**Results and Discussion**

Data from the experiments were analyzed in terms of average sensations of visually and haptically (exploration with the two hands) perceived curvature (average subjective magnitudes and their standard deviations) as a function of the local curve parameter sagitta, or height of the curve (*H*), and as a function of the scale invariant parameter aspect ratio (*H/W*).

The data were then subjected to mathematical modelling and compared across the two sensory modalities to examine whether common characteristics are found [28-30].

*Visual curvature sensing*

Visually perceived curvature was found to increase linearly with the local parameter sagitta, or height (*H*) of the curve. The goodness of the linear fit is satisfactory, as indicated by a linear regression coefficient ($R^2$) of .98. This finding is entirely consistent with previous data [20] showing that sensations of curvature in the visual modality are a linear function of the sagitta. When plotted as a function of the aspect ratio (*H/W*), average visual magnitude of curvature is found to increase with the aspect ratio according to a power law, with an exponent of 3.57 and a correlation coefficient ($R^2$) of .96. The goodness of the power fit is statistically significant ($F(1, 9) = 211.86, p < .001$).

*Tactile curvature sensing*

Haptically perceived curvature by seeing individuals who were blindfolded and could not see the curves, and rely mostly on their vision in everyday life was, like visual curvature, found to increase linearly with the local parameter *H*. The goodness of the linear fit to the tactile data was as for the visual data, with a linear regression coefficient ($R^2$) of .98. When plotted as a function of the aspect ratio (*H/W*), average haptically sensed magnitude of curvature is, like visually perceived curvature, found to increase with the aspect ratio according to a power law, with an exponent of 2.6 and a correlation coefficient (R2) of .90. The goodness of the power fit is statistically significant ($F(1, 9) = 81.91, p < .001$).

Curvature scaled through touch by congenitally blind individuals, who never had visual experience and rely mostly on their other senses in everyday life, was also found to increase linearly with *H*. Analysis of the individual data of each of them shows that the goodness of the linear fits to the data of blind observers is excellent, with regression coefficients ($R^2$) of .95 and .98. When plotted as a function of the aspect ratio (*H/W*), magnitude of curvature haptically sensed by the congenitally blind observers is, again, found to increase with the aspect ratio according to a power law. The goodness of the power fits was statistically significant for both observers ($F(1, 9) = 809.35, p < .001$ and $F(1, 9) = 88.41, p < .001$). The exponents of the power functions are similar (2.97 and 3.19). Whether such statistically robust data can be expected from any congenitally blind patient will largely depend on their ability to perform the psychophysical scaling task and whether other cognitive impairments are diagnosed. Two, otherwise healthy, patients were tested in the experiment as representatives of a larger population of, otherwise healthy, congenitally blind patients.

*The important role of curve symmetry*

In affine geometry, curves derived from circles and ellipses share certain properties, the circle being a particular case of the ellipse. Projective geometry permits, as shown here, generating symmetric curves from ellipses by affinity with concentric circles. Their perception is grounded in biology [31-56] in the sense that most natural objects can be represented in 2D as symmetrically curved shapes with Euclidean properties of ellipses. Also, evolution and geometry generate complexity in similar ways. Evolution drives natural selection, and geometry may capture the logic of this selection and allow expressing it visually, in terms of specific generic properties representing some kind of advantage. Geometry also delivers an account for the logic of evolutionary selection for symmetry, which is found in the shape curves of vein systems and other natural objects such as leaves, cell membranes, and the tunnel systems built by ant populations, just to give a few of many more examples. The topology and geometry of symmetry is controlled by numerical parameters, which act in analogy with a biological organism's DNA. Thus, it is not surprising that symmetry plays an important role in conceptual processes and the design geometry of complex spatial structures, and is abundantly exploited by engineers and architects. The use of the symmetry of curvature, for example, dates back to the dawn of building shelter and vernacular architecture, which relies, by the nature of the materials and construction techniques used, almost entirely on symmetrical curves. In the middle ages, descriptive geometry was used for the planning and execution of building projects for which symmetric curves were the reference model, as in the design of arched hallways and corridors. Studies comparing between visually perceived curvature by experts in geometry (architects and design engineers) and non-experts, using symmetric curves derived from concentric circles by affine projection have shown that their perceived magnitude is determined by a single geometric parameter, the curves' *aspect ratio*. The perceptual responses [57-61] to such curves are independent of both expertise and sensory modality, given that tactile sensing by sighted blindfolded and congenitally blind observers produces the same results. The symmetry of the curves, however, is a critical factor to these geometry-based perceptual responses. The *aspect ratio* relates the height (*sagitta*) to the width of a curve, and in symmetric curves of variable size but constant *aspect ratio* directly taken from concentric circles (no projection by affinity), perceived curvature is also constant, in both vision and touch. This observation is directly linked to the phenomenon of scale-invariance in visual curvature discrimination and the detection and recognition of other shape properties [62-72].

**Conclusions**

Symmetrical curvature is consistently scaled by the human brain on the basis of internal representations. These internal representations are, as the results reviewed herein suggest, statistically invariant, whether they are derived from visual or tactile sensations [73-74]. Physiological data and recent biologically motivated models [75-76] point towards the functional significance of in-built knowledge representations in neural pathways, governing mechanisms through which sensory feedback signals enable the brain to adaptively control and balance motor responses to physical stimuli.

**References**

[1] G. di Pellegrino, E. Ladavas and A. Farnè (1997) Seeing where your hands are. *Nature*, **388**, 730.

[2] A. Farnè and E. Ladavas (2000) Dynamic size-change of hand peripersonal space following tool use. *Neuroreport*, **11**, 1645-1649.

[3] R. Held (2009) Visual-haptic mapping and the origin of crossmodal identity. *Optometry and Vision Science*, **86**, 595-598.

[4] A. Maravita *et al.* (2001) Reaching with a tool extends visual-tactile interactions into far space: evidence from cross-modal extinction. *Neuropsychologia*, **39**, 580-585.

[5] A. Maravita and A. Ikiri (2004) Tools for the body (schema). *Trends in Cognitive Sciences*, **8**, 79-86.

[6] M. Tsuda, Y. Shigemoto-Mogami, S. Koizumi, A. Mizokoshi, S. Kohsaka, M. W. Salter, K. Inoue (2003) P2X4 receptors induced in spinal microglia gate tactile allodynia after nerve injury. *Nature*, **424**, 778-783.

[7] M. S. Graziano and C. G. Cross (1995) The representation of extrapersonal space: a possible role for bimodal, visual-tactile neurons. In: Gazzaniga M. S. (Ed). The cognitive neurosciences. MIT Press: Cambridge, MA, pp. 1054-1057.

[8] C. L. Colby and J. R. Duhamel (1996) Spatial representation for action in parietal cortex. *Cognitive Brain Research*, **5**, 105-115.

[9] Y. Iwamura, A. Iriki, M. Tanaka (1994) Bilateral hand representation in the postcentral somatosensory cortex. *Nature*, **369**, 554-556.

[10] G. Rizzolatti, G. Luppino, M. Matelli (1998) The organisation of the cortical motor system : new concepts. *Electroencephalography and Clinical Neurophysiology*, **106**, 283-296.

[11] P. H. Thakur, P. J. Fitzgerald and S. S. Hsiao (2012) Second-order receptive fields reveal multidigit interactions in area 3B of the macaque monkey. *Journal of Neurophysiology*, **108**, 243-262.

[12] J. J. Gibson (1962) Observations on active touch. *Psychological Review*, **69**, 477-491.

[13] J. J. Gibson (1966) The senses considered as perceptual systems. Houghton Mifflin: Boston, MA.

[14] J. T. Groves (2009) The physical chemistry of membrane curvature. *Nature Chemical Biology*, **5**, 783-784.

[15] N. S. Hatzakis, V. K. Bhatia, J. Larsen, K. L. Madsen, P. Y. Bolinger, A. H. Kunding, J. Castillo, U. Gether, P. Hedegård and D. Stamou (2009) How curved membranes recruit amphipathic helices and protein anchoring motifs. *Nature Chemical Biology*, **5**, 835-841.

[16] R. J. Watt and D. P. Andrews (1982) Contour curvature analysis: hyperacuities in the discrimination of detailed shape. *Vision Research*, **22**, 449-460.

[17] R. J. Watt, R. M. Ward and C. Casco (1987) The detection of deviation from straightness in lines. *Vision Research*, **27**, 1659-1678.

[18] D. H. Foster, D.R. Simmons and M. J. Cook (1993) The cue for contour-curvature discrimination. *Vision Research*, **33**, 329-341.

[19] D. Whitaker and P. V. McGraw (1998) Geometric representation of the mechanisms underlying human curvature detection. *Vision Research*, **38**, 3843-3848.

[20] B. Dresp, C. Silvestri and R. Motro (2007) Which geometric model for the curvature of 2D shape contours? *Spatial Vision*, **20**, 219-265.

[21] S. S. Stevens (1956) The direct estimation of sensory magnitudes. *American Journal of Psychology*, **69**, 1-25.

[22] S. S. Stevens (1975) *Psychophysics*. New York, Wiley.

[23] D. L. Young and C. S. Poon (2001) Soul searching and heart throbbing for biological modeling. *Behavioral and Brain Sciences*, **24**, 1080-1081.

[24] C. S. Poon (1996) Self-tuning optimal regulation of motor output by Hebbian covariance learning. *Neural Networks*, **9**, 1367–1383.

[25] J. M. Foley, N. P. Ribeiro-Filho and J. A. da Silva (2004) Visual perception of extent and the geometry of visual space. *Vision Research*, **44**, 147-156.

[26] C. Q. Howe and D. Purves (2005) Natural scene geometry predicts the perception of angles and line orientation. *Proceedings of the National Academy of Science, USA,* **102**, 1228-1233.

[27] K. A. Stevens (1981) The visual interpretation of surface contours. *Artificial Intelligence*, **17**, 47-73.

[28] Dresp-Langley, B. (2013) Generic Properties of Curvature Sensing through Vision and Touch. *Computational and Mathematical Methods in Medicine*, **634168.**
http://dx.doi.org/10.1155/2013/634168

[29] Dresp-Langley, B. (2015). 2D Geometry Predicts Perceived Visual Curvature in Context-Free Viewing. *Computational Intelligence and Neuroscience*, **708759.**
http://doi.org/10.1155/2015/708759

[30] Dresp-Langley, B. (2016) Affine Geometry, Visual Sensation, and Preference for Symmetry of Things in a Thing. *Symmetry*, **8**, 127. http://www.mdpi.com/2073-8994/8/11/127

[31] Dresp, B., & Grossberg, S. (1997). Contour integration across polarities and spatial gaps: From local contrast filtering to global grouping. *Vision Research*, **37**, 913-924.

[32] Dresp, B. (1997). On illusory contours and their functional significance. *Current Psychology of Cognition*, **16**, 489-518.

[33] Dresp, B. (1998). The effect of practice on the visual detection of near-threshold lines. *Spatial Vision*, **11**, 1-13.

[34] Wehrhahn C., Dresp B. (1998). Detection facilitation by collinear stimuli in humans: Dependence on strength and sign of contrast. *Vision Research*, **38**, 423-428.

[35] Dresp, B. (1998). Area, surface, and contour: Psychophysical correlates of three classes of pictorial completion. *Behavioral and Brain Sciences*, **21**, 755-756.

[36] Dresp, B. (1999). Dynamic characteristics of spatial mechanisms coding contour structures. *Spatial Vision*, **12**, 129-142.

[37] Dresp, B. (1999). The cognitive impenetrability hypothesis: Doomsday for the unity of the cognitive neurosciences? *The Behavioral & Brain Sciences*, **22**, 375-376.

[38] Dresp, B., & Grossberg, S. (1999). Spatial facilitation by colour and luminance edges: boundary, surface, and attention factors. *Vision Research*, **39**, 3431-3443.

[39] Pins, D., Bonnet, C., & Dresp, B. (1999). Response times to brightness illusions of varying subjective magnitude. *Psychonomic Bulletin & Review*, **6**, 437-444.

[40] Dresp, B. (2000). Do positional thresholds define a critical boundary in long-range detection facilitation with co-linear lines? *Spatial Vision*, **13**, 343-357.

[41] Fischer, S., & Dresp, B. (2000). A neural network for long-range contour diffusion by visual cortex. *Lecture Notes in Computer Science*, **1811**, 336-342.

[42] Dresp, B., & Fischer, S. (2001). Asymmetrical contrast effects induced by luminance and colour configurations. *Perception & Psychophysics*, **63**, 1262-1270.

[43] Dresp, B., & Spillmann, L. (2001). The brain may know more than cognitive theory can tell us: a reply to Ted Parks. *Perception*, **30**, 633-636.

[44] Bonnet, C., & Dresp, B. (2001). Investigations of sensory magnitude and perceptual processing with reaction times. *Psychologica*, **25**, 63-86.

[45] Dresp, B. (2002). External regularities and adaptive signal exchanges in the brain. *Behavioral & Brain Sciences*, **24**, 663-664. http://dx.doi.org/10.1017/S0140525X01310087

[46] Tzvetanov, T., & Dresp, B. (2002). Short- and long-range effects in line contrast detection. *Vision Research*, **42**, 2493-2498. http://dx.doi.org/10.1016/S0042-69890200198-0

[47] Dresp, B., Durand, S., & Grossberg, S. (2002). Depth perception from pairs of stimuli with overlapping cues in 2-D displays. *Spatial Vision*, **15**, 255-276. http://dx.doi.org/10.1163/15685680260174038

[48] Dresp, B (2003). Double, double, toil and trouble, fire burn – theory bubble ! Commentary on Steve Lehar's Gestalt Bubble Theory. *Behavioral & Brain Sciences*, **25**, 409-410. http://dx.doi.org/10.1017/S0140525X03230090

[49] Guibal, C, & Dresp, B (2004). Interaction of colour and geometric cues in depth perception: When does « red » mean « near »? *Psychological Research*, **10**, 167-1780. http://dx.doi.org/10.1007/s00426-003-0167-0

[50] Dresp, B., & Langley, O. K. (2005). Long-range spatial integration across contrast signs: a probabilistic mechanism? *Vision Research*, **45**, 275-284. http://dx.doi.org/10.1016/j.visres.2004.08.018

[51] Dresp, B, & Barthaud, JC (2006) Has the brain evolved to answer binding questions or to generate likely hypotheses about the physical world? *Behavioral and Brain Sciences*, **29**, 75−76. http://dx.doi.org/10.1017/S0140525X06279020

[52] Dresp-Langley B, & Quirant, J (2008) Design principles and use of compression structures with tensile integrity. *Recent Patents on Engineering*, **2**, 165-173.

[53] Dresp-Langley, B, & Durup, J (2009) A plastic temporal code for conscious state generation in the brain. *Neural Plasticity*. http://dx.doi.org/10.1155/2009/482696

[54] Dresp-Langley, B (2009) The Ehrenstein illusion. *Scholarpedia*, **4** (10), 53-64. http://dx.doi.org/10.4249/scholarpedia.5364

[55] Silvestri, C, Motro, R, Maurin, B, & Dresp-Langley, B (2010) Visual spatial learning of complex object morphologies through virtual and real-world data. *Design Studies*, **31**, 363−381. http://dx.doi.org/10.1016/j.destud.2010.03.001

[56] Boumenir, Y, Georges, F, Rebillard, G, Valentin, J, Dresp−Langley, B (2010) Wayfinding through an unfamiliar environment. *Perceptual and Motor Skills*, **111**, 1−18. http://dx.doi.org/10.2466/04.22.23.27.PMS.111.6.829-847

[57] Dresp-Langley, B (2014) On "Galileo's visions: Piercing the spheres of the heavens by eye and mind. *Perception*, **43**, 1280-1282. http://dx.doi.org/10.1068/p4311rvw

[58] Dresp-Langley, B, & Durup, J (2012) Does consciousness exist independently of present time and present time independently of consciousness? *Open Journal of Philosophy*, **2**, 45-49. http://dx.doi.org/10.4236/ojpp.2012.21007

[59] Dresp-Langley, B (2012) Why the brain knows more than we do: Non-conscious representations and their role in the construction of conscious experience. *Brain Sciences*, **2**(1), 1-21. http://dx.doi.org/10.3390/brainsci2010001

[60] Dresp-Langley, B, & Reeves, A. (2012) Simultaneous brightness and apparent depth from true colors on grey: Chevreul revisited. *Seeing and Perceiving*, **25**, 597-618. http://dx.doi.org/10.1163/18784763-00002401

[61] Dresp-Langley, B., & Reeves, A. (2014) Effects of saturation and contrast polarity on the figure-ground organization of color on gray. *Frontiers in Psychology*, **5**, 1136.

[62] Boumenir, Y, Vérine, B, Rebillard, G, & Dresp-Langley, B (2014) The relative advantage of tactile 2D route representations for navigation without sight through a complex urban environment. *Terra Haptica*, **4**, 33-44. http://www.psychomot.ups-tlse.fr/Mazella2014.pdf

[63] Spillmann, L, Dresp-Langley, B, & Tseng, CH (2015) Beyond the classic receptive field: The effect of contextual stimuli. *Journal of Vision*, **15**, 7. http://dx.doi.org/10.1167/15.9.7

[64] Dresp-Langley, B., & Grossberg, S. (2016). Neural Computation of Surface Border Ownership and Relative Surface Depth from Ambiguous Contrast Inputs. *Frontiers in Psychology*, **7**, 1102. http://doi.org/10.3389/fpsyg.2016.01102

[65] Silvestri, C, Motro, R, & Dresp-Langley, B (2007) Stevens' Law predicts visual sensations of 2-D curvature. *Perception*, **36** (ECVP Supplement), 38. http://www.perceptionweb.com/ECVP.html

[66] Boumenir, Y, Rebillard, G, Dresp-Langley, B (2011) Brief visual exposure to spatial layout and navigation from memory through a complex urban environment. *Perception*, **40**, (ECVP Supplement), 82. http://www.perceptionweb.com/ECVP.html

[67] Dresp-Langley, B, Reeves, A (2012) Chevreul's laws of contrast and color revisited. *Perception*, **41** (ECVP Supplement), 62. http://www.perceptionweb.com/ECVP.html

[68] Dresp-Langley, B, Reeves, A (2014) The influence of saturation on figure-ground segregation of colored inducers and achromatic backgrounds . *Perception,* **43** (ECVP Supplement). http://journals.sagepub.com/doi/abs/10.1177/03010066140430S101

[69] Dresp-Langley, B (2015) Figure and ground from 2D surfaces with ambiguous border ownership. *Perception*, **44** (ECVP Supplement). http://journals.sagepub.com/doi/full/10.1177/0301006615598674

[70] Reeves, A, Dresp-Langley, B (2017) Perceptual Categories Derived from Reid's "Common Sense" Philosophy. *Frontiers in Psychology,* **8:893**. http://doi.org/10.3389/fpsyg.2017.00893

[71] Dresp-Langley, B, Reeves, A, Grossberg, S (2017) Editorial: Perceptual Grouping—The State of The Art. *Frontiers in Psychol*ogy, **8:67**. http://doi.org/10.3389/fpsyg.2017.00067

[72] Batmaz, A, de Mathelin, M, & Dresp-Langley, B (2016) Effects of Indirect Screen Vision and Tool-Use on the Time and Precision of Object Positioning on Real-World Targets. *Perception*, **45** (ECVP Supplement).

http://journals.sagepub.com/doi/full/10.1177/0301006616671273

[73] Dresp-Langley, B. (2015). Principles of perceptual grouping: implications for image-guided surgery. *Frontiers in Psychology*, **6**, 1565. http://doi.org/10.3389/fpsyg.2015.01565

[74] Batmaz, A. U., de Mathelin, M., & Dresp-Langley, B. (2016). Getting nowhere fast: trade-off between speed and precision in training to execute image-guided hand-tool movements. *BMC Psychology*, *4*, 55. http://doi.org/10.1186/s40359-016-0161-0

[75] Amir, O., Biederman, I. and Hayworth, K.J. (2012). Sensitivity to non-accidental properties across various shape dimensions. *Vision Research,* **62**, 35-43.

[76] Amir, O., Biederman, I., Herald, S.B., Shah, M.P. and Mintz, T.H. (2014). Greater sensitivity to nonaccidental than metric shape properties in preschool children. *Vision Research,* **97**, 83-88.

**Figure captions**

Figure 1

Most objects represented by line contours in the two-dimensional image plane cover a space that roughly corresponds to the shape of an ellipse. The receptive field structures of visual cortical detectors in the primate brain also cover areas which are roughly elliptic. The height-to-with ratio ($h/w$), sometimes also called *aspect ratio*, of 2D shapes is a geometric parameter relative to the visual area covered by a curve.

Figure 2

Curvature selective visual cortical neurons of one and the same coding population respond optimally to deviations from a single straight line on the basis of functionally identified receptive field properties, which include contrast sensitivity and selectivity to local contrast signs (shown here schematically, for illustration). A multitude of such curvature mechanisms operate in parallel in the primates' visual brain and communicate with other sensory systems.

Figure 3

Vertically and horizontal oriented ellipses in the two-dimensional plane can be obtained from concentric circles through a geometric transform called planar projection by affinity. In Cartesian space, an ellipse may be defined as the projected image of two concentric circles. In the two examples given here, images $(x, y) = (b^x, a^y)$ of the principal circle $C_{(0,a)}$ and images $(x, y) = ((a/b) x, y)$ of the secondary circle $C_{(0,b)}$ generate ellipses through planar projection by affinity with the two circles. Upward oriented and downward oriented arcs of eleven such ellipses, derived from concentric circles with varying diameter, were generated.

Figure 4

The psychometric function describing the average data across the two sensory modalities is shown here, with errors bars, to summarize the key findings in a single graph. This representation matches the individual vision and touch data from the different experiments. Subjective magnitudes of visual and tactile curvature sensations, with power fit and and exponential-rise-to-maximum fit as alternative model, are plotted as a function of the *aspect ratio (h/w)* of the curves, a scale invariant geometric property.

**Figures**

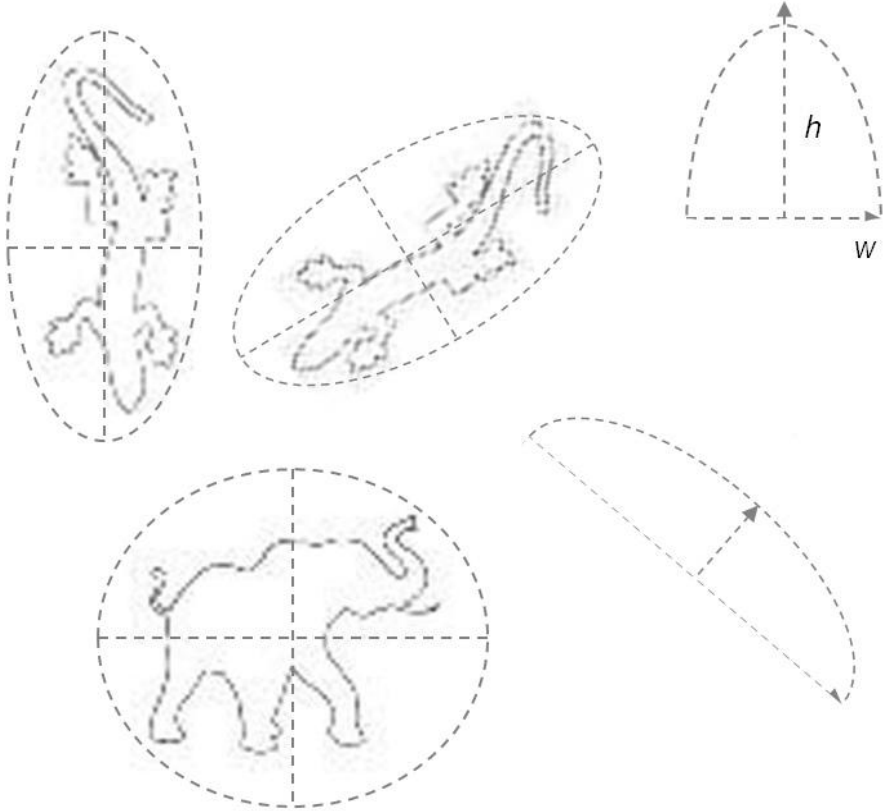

Figure 1

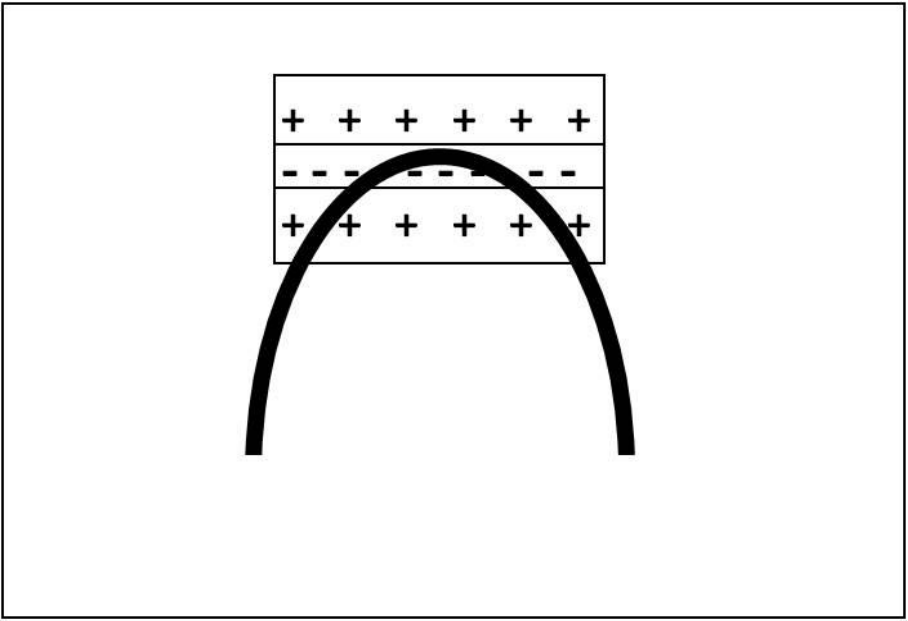

Figure 2

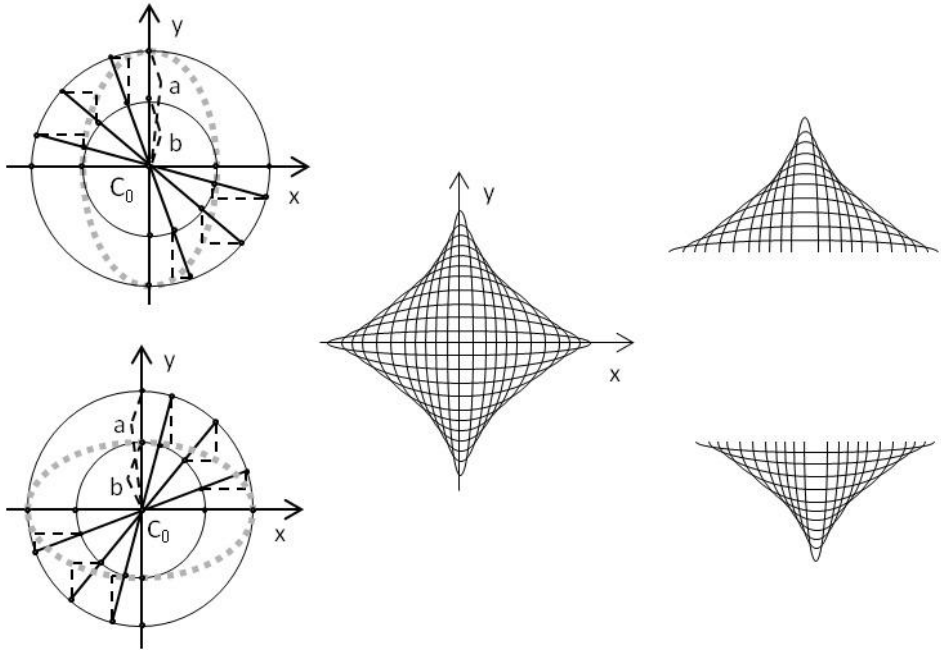

Figure 3

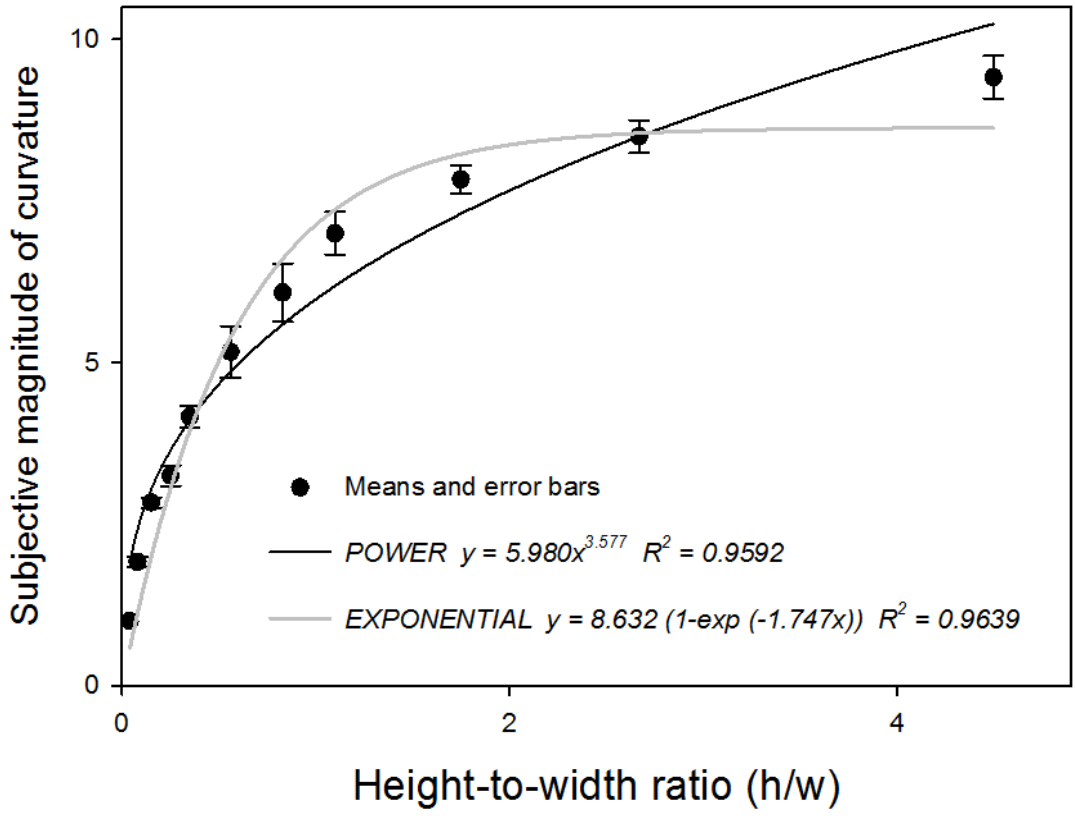

Figure 4